\documentclass[12pt]{article}
\usepackage{enumerate}
\usepackage{amsfonts}
\usepackage{amsmath}
\usepackage{amssymb}
\usepackage{amsthm}

\def\H{\mathcal{H}}

\def\F{\mathfrak{F}}

\def\S{\mathfrak{S}}
\def\C{\mathfrak{C}}
\def\T{\mathfrak{T}}
\def\B{\mathfrak{B}}

\newcommand{\rank}{\mathrm{rank}}
\newcommand{\id}{\mathrm{Id}}
\newcommand{\Tr}{\mathrm{Tr}}

\newcommand{\shs}{\hspace{1pt}}

\newcounter{defin}  \newcounter{lemma} \newcounter{theorem}
\newcounter{property} \newcounter{corol}  \newcounter{remark} \newcounter{example}

\newenvironment{lemma}{\par\refstepcounter{lemma}
     \textbf{Lemma \thelemma.} }{\rm\par}

\newenvironment{property}{\par\refstepcounter{property}
     \textbf{Proposition \theproperty.}\ }{\rm\par}
\newenvironment{corollary}{\par\refstepcounter{corol}
     \textbf{Corollary \thecorol.} }{\rm\par}

\newenvironment{remark}{\par\refstepcounter{remark}
     \textbf{Remark \theremark.}}{\rm\par}

\begin{document}

\title{Energy-constrained diamond norms and their use in quantum information theory}
\author{M.E. Shirokov\footnote{Steklov Mathematical Institute, RAS, Moscow, email:msh@mi.ras.ru}}
\date{}
\maketitle

\begin{abstract}
We consider the family of energy-constrained diamond norms on the set of Hermitian-preserving
linear maps (superoperators) between Banach spaces of trace class operators. We prove that any norm from this family generates the strong (pointwise) convergence on the set of all quantum channels (which is more adequate for describing variations of infinite-dimensional  channels than the diamond norm topology).

We obtain continuity bounds for information characteristics (in particular, classical capacities) of energy-constrained quantum channels (as functions of a channel) with respect to the energy-constrained diamond norms which imply uniform continuity of these characteristics  with respect to the strong convergence topology.
\end{abstract}

\tableofcontents

\section{Introduction}

The diamond-norm distance between two quantum channels is widely used as a  measure of distinguishability between these channels \cite[Ch.9]{Wilde}.
But the topology (convergence) generated by the diamond-norm distance on the set of infinite-dimensional quantum channels is too
strong for analysis of real variations of such  channels. Indeed, for any sequence of unitary operators  $U_n$ strongly converging to the unit operator $I$ but not converging to $I$ in the operator norm  the sequence of channels $\rho\mapsto U_n\rho U^*_n$ does not converge to the identity channel with respect to the diamond-norm distance. In general, the closeness of two quantum channels in the diamond-norm distance means, by Theorem 1 in \cite{Kr&W}, the operator norm closeness of
the corresponding Stinespring isometries. So, if we use the diamond-norm distance then we take into account only such
perturbations of a channel that corresponds to \emph{uniform
deformations} of the Stinespring isometry (i.e. deformations with
small operator norm). As a result, there exist quantum channels with close physical parameters (quantum limited attenuators) having the diamond-norm distance equal to $2$ \cite{W-pc}.

To take into account deformations of the Stinespring
isometry in the strong operator topology one can consider the \emph{strong convergence topology} on the set of all quantum channels defined by the family of seminorms\break $\Phi\mapsto\|\Phi(\rho)\|_1, \rho\in\S(\H_A)$ \cite{AQC}. The convergence of a sequence of channels $\Phi_n$ to a channel $\Phi_0$ in this topology means that
$$
\lim_{n\rightarrow\infty}\Phi_n(\rho)=\Phi_0(\rho)\,\textup{ for all }\rho\in\S(\H_A).
$$
In this paper we show that the strong convergence topology on the set of quantum channels is generated by \emph{any} of
the energy-constrained diamond norms on the set of  Hermitian-preserving
linear maps (superoperators) between Banach spaces of trace class operators (provided the input Hamiltonian satisfies the particular condition).

The energy-constrained diamond norms turn out to be effective tool for quantitative continuity analysis of information characteristics of energy-constrained  quantum channels (as functions of a channel). They can be used instead of the diamond norm in  standard argumentations, in particular, in the Leung-Smith telescopic method used in \cite{L&S} for continuity analysis of capacities of finite-dimensional channels.

We obtain continuity bounds for information characteristics (in particular, classical capacities) of energy-constrained quantum channels  with respect to the energy-constrained diamond norms. They imply, by the above-mentioned correspondence between these norms and the strong convergence, the uniform continuity of these characteristics  with respect to the strong convergence topology.

\section{Preliminaries}

Let $\mathcal{H}$ be a separable infinite-dimensional Hilbert space,
$\mathfrak{B}(\mathcal{H})$ the algebra of all bounded operators with the operator norm $\|\cdot\|$ and $\mathfrak{T}( \mathcal{H})$ the
Banach space of all trace-class
operators in $\mathcal{H}$  with the trace norm $\|\!\cdot\!\|_1$. Let
$\mathfrak{S}(\mathcal{H})$ be  the set of quantum states (positive operators
in $\mathfrak{T}(\mathcal{H})$ with unit trace) \cite{H-SCI,Wilde}.

Denote by $I_{\mathcal{H}}$ the unit operator in a Hilbert space
$\mathcal{H}$ and by $\id_{\mathcal{\H}}$ the identity
transformation of the Banach space $\mathfrak{T}(\mathcal{H})$.\smallskip

If quantum systems $A$ and $B$ are described by Hilbert spaces  $\H_A$ and $\H_B$ then the bipartite system $AB$ is described by the tensor product of these spaces, i.e. $\H_{AB}\doteq\H_A\otimes\H_B$. A state in $\S(\H_{AB})$ is denoted $\rho_{AB}$, its marginal states $\Tr_{\H_B}\rho_{AB}$ and $\Tr_{\H_A}\rho_{AB}$ are denoted respectively $\rho_{A}$ and $\rho_{B}$.
\smallskip

The \emph{von Neumann entropy} $H(\rho)=\mathrm{Tr}\eta(\rho)$ of a
state $\rho\in\mathfrak{S}(\mathcal{H})$, where $\eta(x)=-x\log x$,
is a concave lower semicontinuous function on $\mathfrak{S}(\mathcal{H})$ taking values in $[0,+\infty]$ \cite{L-2,W}.
We will use the binary entropy $h_2(x)=\eta(x)+\eta(1-x)$ and the function
$g(x)=(1+x)h_2\!\left(\frac{x}{1+x}\right)=(x+1)\log(x+1)-x\log x$.\smallskip

The \emph{quantum relative entropy} for two states $\rho$ and
$\sigma$  is defined as
$$
H(\rho\shs\|\shs\sigma)=\sum_i\langle
i|\,\rho\log\rho-\rho\log\sigma\,|i\rangle,
$$
where $\{|i\rangle\}$ is the orthonormal basis of
eigenvectors of the state $\rho$ and it is assumed that
$H(\rho\shs\|\shs\sigma)=+\infty$ if $\,\mathrm{supp}\rho\shs$ is not
contained in $\shs\mathrm{supp}\shs\sigma$ \cite{L-2,W}.
\smallskip

The \emph{quantum mutual information} (QMI) of a  state $\,\rho_{AB}\,$ is defined as
\begin{equation}\label{mi-d}
I(A\!:\!B)_{\rho}=H(\rho_{AB}\shs\Vert\shs\rho_{A}\otimes
\rho_{B})=H(\rho_{A})+H(\rho_{B})-H(\rho_{AB}),
\end{equation}
where the second expression  is valid if $\,H(\rho_{AB})\,$ is finite \cite{L-mi}.
Basic properties of the relative entropy show that $\,\rho\mapsto
I(A\!:\!B)_{\rho}\,$ is a lower semicontinuous function on the set
$\S(\H_{AB})$ taking values in $[0,+\infty]$.

A \emph{quantum channel} $\,\Phi$ from a system $A$ to a system
$B$ is a completely positive trace preserving superoperator
$\mathfrak{T}(\mathcal{H}_A)\rightarrow\mathfrak{T}(\mathcal{H}_B)$,
where $\mathcal{H}_A$ and $\mathcal{H}_B$ are Hilbert spaces
associated with these systems \cite{H-SCI,Wilde}.\smallskip

Denote by $\mathfrak{F}(A,B)$ the set of all quantum channels from  $A$ to
$B$. There are different nonequivalent metrics  on  $\mathfrak{F}(A,B)$. One of them is  induced by the diamond norm
$$
\|\Phi\|_{\diamond}\doteq \sup_{\rho\in\S(\H_{AR})}\|\Phi\otimes \id_R(\rho)\|_1
$$
of a Hermitian-preserving
superoperator $\Phi:\T(\H_A)\rightarrow\T(\H_B)$ \cite{Kit}. The latter  coincides with the norm of complete boundedness of the dual map $\Phi^*:\B(\H_B)\rightarrow\B(\H_A)$ to  $\Phi$ \cite{Paul}.

\section{Some facts about the strong convergence topology on the set of quantum channels}

The strong convergence topology on the set $\F(A,B)$ of quantum channels from $A$ to $B$ is generated by the strong operator
topology on the set of all linear bounded operators from the Banach space
$\T(\H_A)$ into the Banach space $\T(\H_B)$ \cite{R&S}. This topology is studied in detail in \cite{AQC} where it is used for approximation of infinite-dimensional quantum channels by finite-dimensional ones.
Separability of the set $\S(\H_A)$ implies
that the strong convergence topology on the set $\F(A,B)$ is metrisable
(can be defined by some metric). The convergence of a sequence $\{\Phi_n\}$ of channels to a channel $\Phi_0$ in this topology means that
$$
\lim_{n\rightarrow\infty}\Phi_n(\rho)=\Phi_0(\rho)\,\textup{ for all }\rho\in\S(\H_A).
$$

We will use the following simple observations easily proved by using boundedness of the operator norm of all quantum channels.
\smallskip
\begin{lemma}\label{one} \cite{AQC}
\emph{The strong convergence topology on $\,\F(A,B)$  coincides with the topology of uniform convergence on
compact subsets of $\,\S(\H_A)$.}
\end{lemma}
\smallskip
\begin{lemma}\label{two} \cite{AQC}
\emph{The strong convergence of a sequence $\,\{\Phi_n\}\subset\F(A,B)$ to a channel $\,\Phi_0$ implies strong convergence of the sequence $\,\{\Phi_n\otimes \id_R\}$ to the channel $\,\Phi_0\otimes \id_R$, where $R$ is any system.}
\end{lemma}
\smallskip

It is the strong convergence topology  that
makes the set $\F(A,B)$ of all channels  \emph{topologically}
isomorphic to a subset of states of a composite system (generalized
Choi-Jamiolkowski isomorphism).\smallskip

\begin{property}\label{Y-isomorphism} \cite{AQC}
\emph{Let $R\cong A$ and $\,\omega$ be a pure state in
$\,\S(\H_{AR})$ such that $\,\omega_A$ is a full rank state in
$\,\S(\H_{A})$. Then the map
$$
\Phi\mapsto \Phi\otimes
\id_R(\omega)
$$
is a homeomorphism from the set $\,\F(A,B)$ equipped with the strong convergence topology onto the subset $\,\{\rho\in\S(\H_{BR})\,|\,\rho_R=\omega_R\}$.}
\end{property}
\smallskip

If both systems $A$ and $B$ are  infinite-dimensional then
the set $\F(A,B)$ is not compact in the strong convergence topology. Proposition \ref{Y-isomorphism}  implies  the
following compactness criterion for subsets of quantum channels
in this topology.\smallskip

\begin{property}\label{comp-crit} \cite{AQC} \emph{A subset
$\,\F_{0}\subseteq\F(A,B)$ is relatively compact in the strong convergence topology if and only if there exists a full rank
state $\sigma$  in $\,\S(\H_A)$ such that
$\,\{\Phi(\sigma)\}_{\Phi\in\mathfrak{F}_{0}}$ is a relatively compact subset
of $\,\S(\H_B)$. }
\end{property}
\smallskip

Note that the "only if" part of Proposition \ref{comp-crit} is trivial, since
the relative compactness of $\,\F_{0}$ implies
relative compactness of $\,\{\Phi(\sigma)\}_{\Phi\in\F_{0}}$ for arbitrary state $\sigma$
in $\,\S(\H_A)$ by continuity of the map $\Phi\mapsto\Phi(\sigma)$.\smallskip

Proposition \ref{comp-crit} makes it possible to establish  existence of a channel with required properties as a limit point of a sequence of explicitly constructed channels by proving relative compactness of this sequence. For example, by this way one can generalize the famous Petz theorem for two non-faithful (degenerate) infinite rank states starting from the standard version of this theorem (in which both states are assumed faithful) \cite[the Appendix]{RC}.

\section{The energy-constrained diamond norms}

Let $H_A$ be a positive (unbounded) operator in $\H_A$ with dense domain  treated as a Hamiltonian of quantum system $A$. Then $\,\Tr H_A\rho$ is the (mean) energy of a state $\rho\in\S(\H_A)$.\footnote{The value of $\,\Tr H_A\rho$ (finite or infinite) is defined as $\,\sup_n \Tr P_n H_A\rho$, where $P_n$ is the spectral projector of $H_A$ corresponding to the interval $[0,n]$.}\smallskip

Consider the family of norms\footnote{I am grateful to A.Winter who pointed me that formula (\ref{E-sn}) defines a real norm, see the Note Added at the end of the paper.}
\begin{equation}\label{E-sn}
  \|\Phi\|^E_{\diamond}\doteq\sup_{\rho\in\S(\H_{AR}),\Tr H_A\rho_A\leq E}\|\Phi\otimes \id_R(\rho)\|_1,\quad E>E^{A}_0,
\end{equation}
on the set $\mathfrak{L}(A,B)$ of Hermitian-preserving
superoperators from $\T(\H_{A})$ to $\T(\H_{B})$, where $E^{A}_0$ is the infimum of the spectrum of $H_A$ and $R$ is an infinite-dimensional quantum system. They can be  called  \emph{energy-constrained diamond norms (briefly, ECD-norms)}. It is clear that formula (\ref{E-sn}) defines a seminorm on $\mathfrak{L}(A,B)$, the implication $\,\|\Phi\|^E_{\diamond}=0\;\Rightarrow\;\Phi=0\,$ can be  easily shown  by the "convex mixture" arguments from the proof of part A of the following

\smallskip
\begin{property}\label{SCT} \emph{Let $\,\F(A,B)$ be the set of all channels from $\,A$ to $\,B$.}\smallskip

A) \emph{The convergence of a sequence $\,\{\Phi_n\}$ of channels in $\,\F(A,B)$ to a channel $\Phi_0$ with respect to any of the norms in (\ref{E-sn}) implies  the strong convergence of $\,\{\Phi_n\}$ to  $\,\Phi_0$, i.e. for any $E>E^{A}_0$ the following implication holds}
\begin{equation}\label{top-eq}
  \!\!\lim_{n\rightarrow+\infty}\|\Phi_n-\Phi_0\|_{\diamond}^{E}=0\;\; \Rightarrow\;\; \lim_{n\rightarrow+\infty}\Phi_n(\rho)=\Phi_0(\rho)\quad\; \forall \rho\in \S(\H_A).
\end{equation}

B) \emph{If the operator $H_A$ has discrete spectrum $\{E_k\}_{k\geq0}$ of finite multiplicity such that $\,E_k\rightarrow+\infty$ as $k\rightarrow+\infty$ then $\,"\Leftrightarrow"\,$ holds in (\ref{top-eq}) for any $E>E^{A}_0$.}\footnote{It means that the topology generated on the set $\,\F(A,B)$ by any of the norms in (\ref{E-sn})  coincides with the strong convergence topology on $\,\F(A,B)$.}

\end{property}\smallskip

\begin{remark}\label{SCT-r}
The assertions of Proposition \ref{SCT} remain valid for any operator-norm bounded subset of $\mathfrak{L}(A,B)$ (instead of $\F(A,B)$). It follows, in particular, that the strong operator topology on such subsets \emph{is generated by the single norm} $\|\cdot\|_{\diamond}^{E}$ provided that the corresponding operator $H_A$ satisfies the condition of part B.
\end{remark}\smallskip

\emph{Proof.} A) The assumed density of the  domain of $H_A$ in $\H_A$ implies density of the set $\S_0$ of states $\rho$ with finite energy $\,\Tr H_A\rho$ in $\S(\H_A)$. Hence, since the operator norm of all the superoperators $\,\Phi_n-\Phi_0$ is bounded, to prove the implication  in (\ref{top-eq}) it suffices to show that $\,\lim_{n\rightarrow+\infty}\Phi_n(\rho)=\Phi_0(\rho)$ for any $\rho\in\S_0$ provided that  $\lim_{n\rightarrow+\infty}\|\Phi_n-\Phi_0\|_{\diamond}^{E}=0$.

Let $\rho$ be state in $\S_0$ and $\sigma$ a state such that $\Tr H_A\sigma<E$. Then for sufficiently small $p>0$ the energy of the state $\,\rho_p\doteq(1-p)\sigma+p\rho\,$ does not exceed $E$. Hence the left part of (\ref{top-eq}) implies
$$
\lim_{n\rightarrow+\infty}\Phi_n(\rho_p)=\Phi_0(\rho_p)\quad \textrm{and} \quad \lim_{n\rightarrow+\infty}\Phi_n(\sigma)=\Phi_0(\sigma).
$$
It follows that $\Phi_n(\rho)$ tends to $\Phi_0(\rho)$ as $\,n\rightarrow+\infty$. \smallskip

B) By the assumption $\,H_A=\sum_{k=0}^{+\infty} E^A_k|\tau_k\rangle\langle \tau_k|$, where $\{|\tau_k\rangle\}_{k=0}^{+\infty}$ is the orthonormal basis of eigenvectors of $H_A$  corresponding to the nondecreasing sequence $\{E^A_k\}_{k=0}^{+\infty}$ of eigenvalues tending to $+\infty$. Let $P_n=\sum_{k=0}^{n-1}|\tau_k\rangle\langle \tau_k|$ be the projector on the subspace $\H_n$ spanned by the vectors $|\tau_0\rangle,...,|\tau_{n-1}\rangle$. Consider the family of seminorms
\begin{equation}\label{n-sn}
  q_{n}(\Phi)\doteq\sup_{\rho\in\S(\H_n\otimes\H_{R})}\|\Phi\otimes \id_R(\rho)\|_1,\quad n\in\mathbb{N},
\end{equation}
on the set of  Hermitian-preserving
superoperators from $\T(\H_{A})$ to $\T(\H_{B})$. Note that  the system $R$ in (\ref{n-sn}) may be  $n$-dimensional.
Indeed, by convexity of the trace norm the supremum in (\ref{n-sn}) can be taken over pure states $\rho$ in $\S(\H_n\otimes\H_{R})$. Since the marginal state $\rho_R$ of any such pure state $\rho$ has rank $\leq n$, by applying local unitary transformation of the system $R$ we can put all these states into  the set $\S(\H_n\otimes\H'_n)$, where $\H'_n$ is any $n$-dimensional subspace of $\H_R$.\smallskip

Let $\{\Phi_k\}$ be a sequence of channels strongly converging to a channel $\Phi_0$. By Lemmas \ref{one} and \ref{two} this implies that $\,\sup_{\rho\in\C}\|(\Phi_k-\Phi_0)\otimes \id_R(\rho)\|_1$ tends to zero for any compact subset $\C$ of $\S(\H_{AR})$. Since $\dim \H_R=n$, the set $\S(\H_n\otimes\H_R)$ is compact and  we conclude that
\begin{equation}\label{n-sn-lr}
  \lim_{k\rightarrow\infty}q_{n}(\Phi_k-\Phi_0)=0\quad \forall n\in\mathbb{N}.
\end{equation}

Let $E\geq E^A_0$ and $\rho$ be a state in $\S(\H_{AR})$ such that $\Tr H_A\rho_A\leq E$. Then the state $\rho_n=(1-r_n)^{-1}P_n\otimes I_R\;\rho\; P_n\otimes I_R$, where $r_n=\Tr(I_A-P_n)\rho_A$, belongs to the set $\S(\H_n\otimes\H_R)$. By using the inequality
$$
\|(I_A-P_n)\otimes I_R\;\rho\; P_n\otimes I_R\|_1\leq \sqrt{\Tr(I_A-P_n)\otimes I_R\;\rho}=\sqrt{r_n}
$$
easily proved via the operator Cauchy-Schwarz inequality (see the proof of Lemma 11.1 in \cite{H-SCI}) and by noting that $\,\Tr H_A\rho_A\leq E\,$ implies $\,r_n\leq E/E^A_n\,$ we  obtain
$$
\|\rho-\rho_n\|_1\leq 2\|(I_A-P_n)\otimes I_R\;\rho\; P_n\otimes I_R\|_1+2r_n\leq 4\sqrt{r_n}\leq 4\sqrt{E/E^A_n}.
$$
It follows that $\,\|\Phi_k-\Phi_0\|_{\diamond}^{E}\leq q_{n}(\Phi_k-\Phi_0)+8\sqrt{E/E^A_n}$. Since $E^A_n\rightarrow+\infty$ as $n\rightarrow+\infty$, this inequality and (\ref{n-sn-lr}) show that $\,\lim_{k\rightarrow\infty}\|\Phi_k-\Phi_0\|_{\diamond}^{E}=0$. $\square$

\section{Continuity bounds for information characteristics of quantum channels with respect to the ECD-norms}

In this section we show that the ECD-norms can be effectively used for quantitative continuity analysis of information characteristics of energy-constrained  quantum channels (as functions of a channel). They allow (due to Proposition \ref{SCT}) to prove uniform continuity of these characteristics with respect to the strong convergence of quantum channels.\smallskip

In what follows we will consider quantum channels  between given infinite-dimensional systems $A$ and $B$. We will assume that the Hamiltonians $H_A$ and $H_B$ of these systems are densely defined positive operators and that
\begin{equation}\label{H-cond}
  \Tr e^{-\lambda H_{B}}<+\infty\quad\textrm{for all}\;\lambda>0.
\end{equation}
To formulate our  main results introduce the function
\begin{equation}\label{F-def}
F_{H_B}(E)\doteq\sup_{\Tr H_B\rho\leq E}H(\rho),\quad E\geq E^{\!B}_0,
\end{equation}
where $E^{\!B}_0$ is the minimal eigenvalue of $H_B$.
Properties of this function are described in Proposition 1 in \cite{EC}, where it is shown, in particular, that
\begin{equation}\label{F-exp}
F_{H_B}(E)=\lambda(E)E+\log\Tr e^{-\lambda(E) H_{\!B}}=o\shs(E)\quad\textup{as}\quad E\rightarrow+\infty,
\end{equation}
where $\lambda(E)$ is determined by the equality $\Tr H_B e^{-\lambda(E) H_B}=E\Tr e^{-\lambda(E) H_B}$, provided that condition (\ref{H-cond}) holds.

It is well known that condition (\ref{H-cond}) implies  continuity of the von Neumann entropy on the subset of $\S(\H_B)$ determined by the inequality $\Tr H_B\rho\leq E$ for any $E\geq E^{\!B}_0$ and  attainability of the supremum in (\ref{F-def})  at the \emph{Gibbs state} $\gamma_B(E)\doteq e^{-\lambda(E) H_B}/\Tr e^{-\lambda(E) H_B}$ \cite{W}. So, we have $F_{H_B}(E)=H(\gamma_B(E))$ for any $E> E^{\!B}_0$. \smallskip

The function $F_{H_B}$ is increasing and concave on $[E^{\!B}_0,+\infty)$. Denote by $\widehat{F}_{H_B}$ any upper bound for $F_{H_B}$ defined on $[0,+\infty)$ possessing the properties
\begin{equation}\label{F-prop-1}
\widehat{F}_{H_B}(E)> 0,\quad \widehat{F}_{H_B}^{\shs\prime}(E)>0,\quad \widehat{F}_{H_B}^{\shs\prime\prime}(E)< 0\quad\textrm{for all }\; E>0
\end{equation}
and
\begin{equation}\label{F-prop-2}
\widehat{F}_{H_B}(E)=o\shs(E)\quad\textrm{as}\quad E\rightarrow+\infty.
\end{equation}
At least one such function $\widehat{F}_{H_B}$ always exists. It follows from (\ref{F-exp}) that  one can use $F_{H_B}(E+E^{\!B}_0)$ in  the role of $\widehat{F}_{H_B}(E)$. \smallskip

If $B$ is the $\,\ell$-mode quantum oscillator with the Hamiltonian
\begin{equation*}
H_B=\sum_{i=1}^{\ell}\hbar\shs\omega_i \left (a^{+}_ia_i+\textstyle\frac{1}{2}I_B\right),
\end{equation*}
where $\,a_i\,$ and $\,a^{+}_i\,$ are the annihilation and creation operators and $\,\omega_i\,$ is the frequency of the $i$-th oscillator \cite[Ch.12]{H-SCI} then one can show that\footnote{We use the natural logarithm.}
\begin{equation}\label{bF-ub}
F_{H_B}(E)\leq \widehat{F}_{\ell,\omega}(E)\doteq\ell\log \frac{E+E_0}{\ell E_*}+\ell,
\end{equation}
where $E_0\doteq\frac{1}{2}\sum_{i=1}^{\ell}\hbar\omega_i$ and $E_*=\left[\prod_{i=1}^{\ell}\hbar\omega_i\right]^{1/\ell}$ and that  upper bound (\ref{bF-ub}) is $\varepsilon$-sharp for large $E$ \cite{CHI}. It is easy to see that the function $\widehat{F}_{\ell,\omega}$ possesses properties (\ref{F-prop-1}) and (\ref{F-prop-2}). So, it can be used in the role of $\widehat{F}_{H_B}$ in this case.

\subsection{The output Holevo quantity of a channel}

In this subsection we obtain continuity bound for the function $\Phi\mapsto\chi(\Phi(\mu))$ with respect to the ECD-norm, where
$\chi(\Phi(\mu))$ is the Holevo quantity of the image of a given (discrete or continuous) ensemble $\mu$ of input states under the channel $\Phi$.

Discrete  ensemble $\mu=\{p_i,\rho_i\}$ is a finite or countable collection $\{\rho_i\}$ of quantum states  with the corresponding probability distribution $\{p_i\}$. The  Holevo quantity of $\mu$ is defined as
$$
\chi\left(\mu\right)\doteq \sum_{i} p_i H(\rho_i\|\bar{\rho}(\mu))=H(\bar{\rho}(\mu))-\sum_{i} p_i H(\rho_i),
$$
where $\,\bar{\rho}(\mu)=\sum_{i} p_i\rho_i$ is the average state of $\mu$ and the second formula is valid if $H(\bar{\rho}(\mu))<+\infty$. This quantity gives the upper bound for classical information obtained by recognizing states of the ensemble by quantum measurements \cite{H-73}. It plays important role in analysis of information properties of quantum systems and channels \cite{H-SCI,Wilde}.\smallskip

Generalized (continuous) ensembles of quantum states are  defined as
Borel probability measures on the set of quantum states \cite{H-SCI,H-Sh-2}.  The average state of a generalized
ensemble $\mu$ is the barycenter of
$\mu $ defined by the Bochner integral
\begin{equation*}
\bar{\rho}(\mu )=\int\rho \mu (d\rho ).
\end{equation*}

The Holevo quantity of a
generalized ensemble $\mu$ is defined as
\begin{equation*}
\chi(\mu)=\int H(\rho\shs \|\shs \bar{\rho}(\mu))\mu (d\rho )=H(\bar{\rho}(\mu
))-\int H(\rho)\mu (d\rho ),  
\end{equation*}%
where the second formula is valid under the condition $H(\bar{\rho}(\mu))<+\infty$ \cite{H-Sh-2}.\smallskip

The average energy of $\,\mu\,$ is given by the formula
$$
E(\mu)=\Tr H\bar{\rho}(\mu)=\int\Tr H\rho\,\mu(d\rho),
$$
where $H$ is the Hamiltonian of the system (the integral is well defined, since the function $\rho\mapsto\Tr H\rho$ is lower semicontinuous). For a discrete ensemble  $\mu=\{p_i,\rho_i\}$ we have  $\,E(\mu)=\sum_{i} p_i\Tr H\rho_i$.\smallskip

For an ensemble $\mu$ of states in $\S(\H_A)$ its image $\Phi(\mu)$
under a quantum channel $\,\Phi :A\rightarrow B\,$ is defined as the
ensemble corresponding to the measure $\mu
\circ \Phi ^{-1}$ on $\mathfrak{S}(\mathcal{H}_{B})$, i.e. $\,\Phi (\mu )[\S_{\!B}]=\mu[\Phi ^{-1}(\S_{\!B})]\,$
for any Borel subset $\S_{\!B}$ of $\mathfrak{S}(\mathcal{H}_{B})$,
where $\Phi ^{-1}(\S_{\!B})$ is the pre-image of $\S_{\!B}$ under the map $\Phi $.
If $\mu =\{p _{i},\rho _{i}\}$ then this definition implies $\Phi (\mu)=\{p _{i},\Phi(\rho_{i})\}$.\smallskip

\begin{property}\label{CHI-CB} \emph{Let $\,\mu$  be a generalized ensemble of states in $\S(\H_A)$ with finite average energy $E(\mu)$. Let $\,\Phi$ and $\,\Psi$ be  channels from  $A$ to $B$ such that $\,\Tr H_B\Phi(\bar{\rho}(\mu)), \Tr H_B\Psi(\bar{\rho}(\mu))\leq E$ and  $\,\frac{1}{2}\shs \|\Phi-\Psi\|^{E(\mu)}_{\diamond}\leq\varepsilon$. If the Hamiltonian $H_B$ satisfies conditions (\ref{H-cond})  then
\begin{equation}\label{CHI-CB+}
\left|\chi(\Phi(\mu))-\chi(\Psi(\mu))\right|\leq\varepsilon(2t+r_{\!\varepsilon}(t))\widehat{F}_{H_B}\!\!\left(\frac{E}{\varepsilon t}\right)
+2g(\varepsilon r_{\!\varepsilon}(t))+2h_2(\varepsilon t)
\end{equation}
for any $\,t\in(0,\frac{1}{2\varepsilon}]$, where $\,\widehat{F}_{H_B}(E)$ is any upper bound for the function $F_{H_B}(E)$ (defined in (\ref{F-def})) with properties  (\ref{F-prop-1}) and (\ref{F-prop-2}), in particular $\,\widehat{F}_{H_B}(E)=F_{H_B}(E+E^{\!B}_0)$ and $\,r_{\!\varepsilon}(t)=(1+t/2)/(1-\varepsilon t)$.}\footnote{The functions $h_2(x)$ and $g(x)$ are defined in Section 2.}
\smallskip

\emph{If $\,B$ is the $\ell$-mode quantum oscillator  then
\begin{equation}\label{CHI-CB++}
\begin{array}{rl}
\left|\chi(\Phi(\mu))-\chi(\Psi(\mu))\right| \,\leq &\!\!\varepsilon(2t+r_{\!\varepsilon}(t))\left[\displaystyle
\widehat{F}_{\ell,\omega}(E)-\ell\log(\varepsilon t)\right]\\\\
\,+ &\!\! 2g(\varepsilon r_{\!\varepsilon}(t))+2h_2(\varepsilon t)
\end{array}
\end{equation}
for any $\,t\in(0,\frac{1}{2\varepsilon}]$, where $\,\widehat{F}_{\ell,\omega}(E)$ is defined in (\ref{bF-ub}). Continuity bound (\ref{CHI-CB++}) with optimal $\,t$ is  tight for large $E$.}
\end{property}\smallskip

\begin{remark}\label{omi-r} Condition (\ref{F-prop-2}) implies that $\,\displaystyle \lim_{x\rightarrow+0}x\widehat{F}_{H_B}\!(E/x)=0\,$. Hence, the right hand side of (\ref{CHI-CB+}) tends to zero as $\,\varepsilon\rightarrow0$.
\end{remark}\smallskip

\begin{remark}\label{omi-r+} It is easy to see that the right hand side of (\ref{CHI-CB+}) attains minimum at some \emph{optimal} $\,t=t(E, \varepsilon)$. It is this minimum that gives proper upper bound for $\,\left|\chi(\Phi(\mu))-\chi(\Psi(\mu))\right|$.
\end{remark}\medskip

\emph{Proof.} Let  $\mu=\{p_i,\rho_i\}_{i=1}^m$ be a discrete ensemble of $m\leq+\infty$ states and
$\hat{\rho}_{AC}=\sum_{i=1}^m p_i\rho_i\otimes |i\rangle\langle i|$ the corresponding $cq$-state (here $\,\{|i\rangle\}_{i=1}^m$ is an orthonormal basis in a $m$-dimensional Hilbert space $\H_C$). Then $\Tr H_A\hat{\rho}_{A}=E(\mu)$ and hence
\begin{equation}\label{l-ineq}
\|\Phi\otimes\id_C(\hat{\rho}_{AC})-\Psi\otimes\id_C(\hat{\rho}_{AC})\|_1\leq \|\Phi-\Psi\|^{E(\mu)}_{\diamond}.
\end{equation}
So, in this case the assertions of the proposition follow from Proposition 7 in \cite{CHI}, since the left hand side of (\ref{l-ineq}) coincides with
$$
2D_0(\{p_i,\Phi(\rho_i)\},\{p_i,\Psi(\rho_i)\})\doteq\sum_{i=1}^m p_i \|\Phi(\rho_i)-\Psi(\rho_i)\|_1.
$$

Let $\mu$ be an arbitrary generalized ensemble. The construction from the proof of Lemma 1 in \cite{H-Sh-2} gives the sequence $\{\mu_n\}$ of discrete ensembles weakly\footnote{The weak convergence of a sequence $\{\mu_n\}$  to a measure $\mu_0$ means that
$\,\lim\limits_{n\rightarrow\infty}\int f(\rho)\mu_n(d\rho)=\int f(\rho)\mu_0(d\rho)\,$
for any continuous bounded function $f$ on $\,\S(\H)$ \cite{Bil}.} converging to $\mu$ such that $\bar{\rho}(\mu_n)=\bar{\rho}(\mu)$ for all $n$. Since the assumption $\,\Tr H_B\Phi(\bar{\rho}(\mu)), \Tr H_B\Psi(\bar{\rho}(\mu))\leq E\,$ implies $\,H(\Phi(\bar{\rho}(\mu))), H(\Psi(\bar{\rho}(\mu)))<+\infty$, Corollary 1 in \cite{AQC} shows that
$$
\lim_{n\rightarrow\infty}\chi(\Phi(\mu_n))=\chi(\Phi(\mu))\quad\textup{and}\quad\lim_{n\rightarrow\infty}\chi(\Psi(\mu_n))=\chi(\Psi(\mu)).
$$
So, the validity of inequality (\ref{CHI-CB+}) for the ensemble $\mu$ follows from the validity of this inequality for all the discrete ensembles $\mu_n$ proved before. Inequality (\ref{CHI-CB++}) is a direct corollary of (\ref{CHI-CB+}).

The  tightness of continuity bound (\ref{CHI-CB++}) follows from the tightness of  continuity bound (\ref{feaf-cb-1}) for the Holevo capacity in this case (obtained from (\ref{CHI-CB++})).

\smallskip

By Remark \ref{omi-r} Propositons \ref{SCT} and \ref{CHI-CB} imply the following \smallskip

\begin{corollary}\label{CHI-CB-c} \emph{If the Hamiltonians $H_A$ and $H_B$ satisfy, respectively, the condition of Proposition \ref{SCT}B and condition (\ref{H-cond}) then for any generalized ensemble $\,\mu$ of states in $\S(\H_A)$ with finite average energy the function $\,\Phi\mapsto \chi(\Phi(\mu))\,$ is uniformly continuous
on the sets
$$
\F_{\mu,E} \doteq \left\{\shs\Phi\in\F(A,B)\,|\,\Tr H_B\Phi(\bar{\rho}(\mu))\leq E\shs\right\}, \quad E>E^{\!B}_0,
$$
with respect to the strong convergence topology.}
\end{corollary}\smallskip

\begin{remark}\label{CHI-CB-r}
The uniform continuity with respect to the strong convergence topology means  uniform continuity with respect to any metric generating this topology, in particular, with respect to any of the ECD-norms.
\end{remark}

\subsection{QMI at the output of n copies of a local channel}

The following proposition is a corollary of Proposition 3B in \cite{CHI} proved by the Leung-Smith telescopic trick used in \cite{L&S} and  Winter's technique from \cite{W-CB}. It gives tight continuity bounds for the function $\Phi\mapsto I(B^n\!:\!C)_{\Phi^{\otimes n}\otimes\id_{C}(\rho)}$ for any given $n$ and a state $\rho\in\S(\H^{\otimes n}_{A}\otimes\H_{C})$ with respect to the ECD-norm provided that the marginal states $\rho_{A_1},...,\rho_{A_n}$ have finite energy.\smallskip

\begin{property}\label{omi} \emph{Let $\,\Phi$ and $\,\Psi$ be  channels from  $A$ to $B$, $C$ be any system and $\shs\rho$  a state in $\,\S(\H^{\otimes n}_{A}\otimes\H_{C})$ such that $\,E_A\doteq\max_{1\leq k\leq n}\{\Tr H_A\rho_{A_k}\}$ is finite. Let $\,\frac{1}{2}\shs \|\Phi-\Psi\|^{E_A}_{\diamond}\leq\varepsilon$, $\,\Tr H_B\Phi(\rho_{A_k}), \Tr H_B\Psi(\rho_{A_k})\leq E_k$ for $k=\overline{1,n}$ and
$$
\;\Delta^n(\Phi,\Psi,\rho)\doteq\left|I(B^n\!:\!C)_{\Phi^{\otimes n}\otimes\id_{C}(\rho)}-I(B^n\!:\!C)_{\Psi^{\otimes n}\otimes\id_{C}(\rho)}\right|.
$$
If the Hamiltonian $H_B$ satisfies conditions (\ref{H-cond})  then
\begin{equation}\label{CBn-2}
\Delta^n(\Phi,\Psi,\rho)\leq 2n\varepsilon(2t+r_{\!\varepsilon}(t))\widehat{F}_{H_B}\!\!\left(\frac{E}{\varepsilon t}\right)
+2ng(\varepsilon r_{\!\varepsilon}(t))+4nh_2(\varepsilon t)
\end{equation}
for any $\,t\in(0,\frac{1}{2\varepsilon}]$, where $\; E=n^{-1}\sum_{k=1}^nE_k$,  $\,\widehat{F}_{H_B}(E)$ is any upper bound for the function $F_{H_B}(E)$ (defined in (\ref{F-def})) with properties  (\ref{F-prop-1}) and (\ref{F-prop-2}), in particular, $\,\widehat{F}_{H_B}(E)=F_{H_B}(E+E_0)$ and  $\,r_{\!\varepsilon}(t)=(1+t/2)/(1-\varepsilon t)$.}\footnote{The functions $h_2(x)$ and $g(x)$ are defined in Section 2.}\smallskip

\emph{If $\,B$ is the $\ell$-mode quantum oscillator then
\begin{equation}\label{CBn-2+}
\begin{array}{c}
\Delta^n(\Phi,\Psi,\rho)\leq 2n\varepsilon(2t+r_{\!\varepsilon}(t))\!\left[\displaystyle
\widehat{F}_{\ell,\omega}(E)-\ell\log(\varepsilon t)\right]\\\\
+\,2ng(\varepsilon r_{\!\varepsilon}(t))+4nh_2(\varepsilon t),
\end{array}
\end{equation}
where $\widehat{F}_{\ell,\omega}(E)$ is defined in (\ref{bF-ub}). Continuity bound (\ref{CBn-2+}) with optimal $\,t$ is tight for large $E$ (for any given $n$).}
\end{property}\medskip

\begin{remark}\label{qcmi}
All the assertions of Proposition \ref{omi} remain valid for the quantum conditional mutual information (see Proposition 3B in \cite{CHI}).\smallskip
\end{remark}

Since condition (\ref{F-prop-2}) implies $\,\displaystyle \lim_{x\rightarrow+0}x\widehat{F}_{H_B}\!(E/x)=0\,$,  the right hand side of (\ref{CBn-2}) tends to zero as $\varepsilon\rightarrow0$. Hence  Propositons \ref{SCT} and \ref{omi} imply the following \smallskip

\begin{corollary}\label{omi-c} \emph{If the Hamiltonians $H_A$ and $H_B$ satisfy, respectively, the condition of Proposition \ref{SCT}B and condition (\ref{H-cond}) then for any $n\in\mathbb{N}$ and any state $\rho$ in $\,\S(\H^{\otimes n}_{A}\otimes\H_{C})$ such that $\,\max_{1\leq k\leq n}\{\Tr H_A\rho_{A_k}\}<+\infty$  the function $\Phi\mapsto I(B^n\!:\!C)_{\Phi^{\otimes n}\otimes\id_{C}(\rho)}$ is uniformly continuous\footnote{see Remark \ref{CHI-CB-r}.}
on the sets
$$
\F_{\rho,E} \doteq \left\{\shs\Phi\in\F(A,B)\,\left|\,\frac{1}{n}\sum_{k=1}^n \Tr H_B\Phi(\rho_{A_k})\leq E\shs\right.\right\}, \quad E>E^{\!B}_0,
$$
with respect to the strong convergence topology.}
\end{corollary}

\subsection{Classical capacities of infinite-dimensional energy-constrained channels}

When we consider transmission of classical information over infinite
dimensional quantum channels we have to impose the energy constraint on states
used for coding information.  For a single channel $\Phi:A\rightarrow B$ the energy constraint is
determined by the linear inequality
\begin{equation*}
\mathrm{Tr} H_A\rho\le E,
\end{equation*}
where $H_A$ is the Hamiltonian of
the input system $A$. For $n$ copies of this channel the energy constraint is given by the inequality
\begin{equation}\label{constraint}
\mathrm{Tr}H_{A^n}\rho^{(n)}\leq nE,
\end{equation}
where $\rho^{(n)}$ is a state of the system $A^n$ ($n$ copies of $A$) and
\begin{equation}\label{f-n}
H_{A^n}=H_A\otimes \dots \otimes I+\dots +I\otimes \dots \otimes H_A
\end{equation}
is the Hamiltonian of the system $A^n$ \cite{H-SCI,H-c-w-c,Wilde+}.

An operational definition of the classical capacity of an infinite-dimensional  energy-constrained  quantum
channel can be found in \cite{H-c-w-c}. If only  nonentangled input encoding is used then
the ultimate rate of transmission of classical information trough the channel $\Phi$ with the constraint
(\ref{constraint}) on mean energy of a code is determined by the Holevo capacity
\begin{equation}
C_{\chi}(\Phi,H_A,E)=\sup_{\mathrm{Tr}H_A\bar{\rho}\leq E}\chi(\{p_i,\Phi(\rho_i)\}),\quad \bar{\rho}=\sum_i p_i\rho_i,
\label{chi-cap}
\end{equation}
(the supremum is over all input ensembles $\{p_i,\rho_i\}$ such that $\mathrm{Tr}H_A\bar{\rho}\leq E$).
By the Holevo-Schumacher-Westmoreland theorem adapted to constrained
channels (\cite[Proposition 3]{H-c-w-c}), the classical capacity of
the channel $\Phi $ with constraint (\ref{constraint}) is given by
the following regularized expression
\begin{equation*}
C(\Phi ,H_A,E)=\lim_{n\rightarrow +\infty }n^{-1}C_{\chi}(\Phi
^{\otimes n},H_{A^n},nE),
\end{equation*}
where $H_{A^n}$ is defined in (\ref{f-n}). If $C_{\chi}(\Phi
^{\otimes n},H_{A^n},nE)=nC_{\chi}(\Phi,H_A,E)$ for all $n$ then
\begin{equation}\label{additiv}
C(\Phi,H_A,E)=C_{\chi}(\Phi,H_A,E),
\end{equation}
i.e.  the classical capacity of the channel $\Phi$  coincides with
its Holevo capacity. Note that (\ref{additiv}) holds for many
infinite-dimensional channels \cite{H-SCI}.  Recently it was shown that (\ref{additiv}) holds
if $\Phi$ is a  gauge covariant or contravariant Gaussian channel and
$H_A=\sum_{ij}\epsilon_{ij}a^{\dagger}_ia_j$ is a gauge invariant\footnote{The gauge invariance condition for $H_A$ can be replaced by the  condition (18) in \cite{H-L}.}
Hamiltonian  (here $[\epsilon_{ij}]$
is a positive matrix) \cite{GHP,H-L}.

The following proposition presents estimates for differences between the Holevo capacities and between the classical capacities of channels $\Phi$ and $\Psi$ with \emph{finite energy amplification factors for given input energy}, i.e. such that
\begin{equation}\label{feaf}
 \sup_{\Tr H_A\rho\leq E} H_B\Phi(\rho)\leq kE\quad\textrm{and}\quad \sup_{\Tr H_A\rho\leq E} H_B\Psi(\rho)\leq kE
\end{equation}
for  given $E\geq E^{A}_0$ and finite $k=k(E)$. Note that any channels produced in a physical experiment satisfy condition (\ref{feaf}). \smallskip

\begin{property}\label{feaf-cb} \emph{Let $\,\Phi$ and $\,\Psi$ be quantum channels from $A$ to $B$ satisfying condition (\ref{feaf}) and   $\,\frac{1}{2}\shs \|\Phi-\Psi\|^{E}_{\diamond}\leq\varepsilon$. If the Hamiltonian $H_B$ satisfies conditions (\ref{H-cond})  then
\begin{equation}\label{feaf-cb-1}
\begin{array}{rl}
|C_{\chi}(\Phi,H_A,E)-C_{\chi}(\Psi,H_A,E)|\, \leq & \!\!\varepsilon(2t+r_{\!\varepsilon}(t))\widehat{F}_{H_B}\!\!\left(\frac{kE}{\varepsilon t}\right)\\\\
\,+ & \!\!2g(\varepsilon r_{\!\varepsilon}(t))+2h_2(\varepsilon t)
\end{array}
\end{equation}
and
\begin{equation}\label{feaf-cb-2}
\begin{array}{rl}
|C(\Phi,H_A,E)-C(\Psi,H_A,E)|\, \leq & \!\! 2\varepsilon(2t+r_{\!\varepsilon}(t))\widehat{F}_{H_B}\!\!\left(\frac{kE}{\varepsilon t}\right)\\\\
\,+ & \!\!2g(\varepsilon r_{\!\varepsilon}(t))+4h_2(\varepsilon t)
\end{array}
\end{equation}
for any $\,t\in(0,\frac{1}{2\varepsilon}]$, where $\,r_{\!\varepsilon}(t)=(1+t/2)/(1-\varepsilon t)$ and  $\,\widehat{F}_{H_B}(E)$ is any upper bound for the function $F_{H_B}(E)$ (defined in (\ref{F-def})) with properties  (\ref{F-prop-1}) and (\ref{F-prop-2}),  in particular, $\,\widehat{F}_{H_B}(E)=F_{H_B}(E+E^{\!B}_0)$.}\footnote{The functions $h_2(x)$ and $g(x)$ are defined in Section 2.}\smallskip

\emph{If $\,B$ is the $\ell$-mode quantum oscillator then the right hand sides of (\ref{feaf-cb-1}) and (\ref{feaf-cb-2}) can be rewritten, respectively, as follows
\begin{equation*}
\varepsilon(2t+r_{\!\varepsilon}(t))\left[\displaystyle
\widehat{F}_{\ell,\omega}(kE)-\ell\log(\varepsilon t)\right]+2g(\varepsilon r_{\!\varepsilon}(t))+2h_2(\varepsilon t)
\end{equation*}
and
\begin{equation*}
2\varepsilon(2t+r_{\!\varepsilon}(t))\!\left[\displaystyle
\widehat{F}_{\ell,\omega}(kE)-\ell\log(\varepsilon t)\right]+2g(\varepsilon r_{\!\varepsilon}(t))+4h_2(\varepsilon t),
\end{equation*}
where $\widehat{F}_{\ell,\omega}(E)$ is defined in (\ref{bF-ub}). In this case continuity bound (\ref{feaf-cb-1}) with optimal $\;t\,$ is tight for large $E$ while  continuity bound (\ref{feaf-cb-2}) is close-to-tight (up to the factor $2$ in the main term).}
\end{property}\medskip

\emph{Proof.} Inequality (\ref{feaf-cb-1}) follows from definition (\ref{chi-cap}) and Proposition \ref{CHI-CB}.

To prove inequality (\ref{feaf-cb-2}) note that
$$
C_{\chi}(\Phi^{\otimes n},H_{A^n},nE)=\sup\chi(\{p_i,\Phi^{\otimes n}(\rho_i)\}),
$$
where the supremum is over all ensembles  $\{p_i,\rho_i\}$ of states in $\S(\H^{\otimes n}_A)$ with the average state $\bar{\rho}$ such that $\Tr H_A\bar{\rho}_{A_j}\leq E$ for all $j=\overline{1,n}$. This can be easily shown by using symmetry arguments and the following well known property of the Holevo quantity:
$$
\frac{1}{n}\sum_{j=1}^n\chi\left(\{q^j_i,\sigma^j_i\}_i\right)\leq \chi\left(\left\{\frac{q^j_i}{n},\sigma^j_i\right\}_{ij}\right)
$$
for any collection $\,\{q^1_i,\sigma^1_i\},...,\{q^n_i,\sigma^n_i\}\,$ of discrete ensembles. \smallskip

Since condition (\ref{feaf}) implies
\begin{equation*}
  \Tr H_B\Phi(\bar{\rho}_{A_j})\leq kE\quad \textrm{and} \quad \Tr H_B\Psi(\bar{\rho}_{A_j})\leq kE,\quad j=\overline{1,n},
\end{equation*}
for any ensemble  $\{p_i,\rho_i\}$ satisfying the above condition, continuity bound (\ref{feaf-cb-2}) is obtained by using the representations
$$
\chi(\{p_i,\Lambda^{\!\otimes n}(\rho_i)\})=I(B^n\!:\!C)_{\Lambda^{\!\otimes n}\otimes\id_C(\hat{\rho})},\; \Lambda=\Phi,\Psi,\;\textrm{ where }\,\hat{\rho}_{AC}=\sum_{i}p_i\rho_i\otimes |i\rangle\langle i|,
$$
Proposition \ref{omi} and the corresponding analog of Lemma 12 in \cite{L&S}.

If $\,B$ is the $\ell$-mode quantum oscillator and $\,\widehat{F}_{H_B}=\widehat{F}_{\ell,\omega}$ then
we can estimate the right hand sides of (\ref{feaf-cb-1}) and (\ref{feaf-cb-2}) from above by using the inequality
$\,\widehat{F}_{\ell,\omega}(E/x)\leq \widehat{F}_{\ell,\omega}(E)-\ell\log x\,$ valid for any positive $E$ and $x\leq1$.

The  tightness of  continuity bound (\ref{feaf-cb-1})  can be shown assuming that $\Phi$ is the identity channel from the $\ell$-mode  quantum oscillator $A$ to $B=A$ and $\Psi$ is the completely depolarizing channel with the vacuum output state. These channels
satisfy condition (\ref{feaf}) with $k=1$ and
\begin{equation}\label{C-dif}
|C_{\chi}(\Phi,H_A,E)-C_{\chi}(\Psi,H_A,E)| =\sup\limits_{\Tr H_A\rho\leq E}H(\rho)= F_{H_A}(E).
\end{equation}
By Lemma 2 in \cite{CHI} in the case $\,\widehat{F}_{H_B}=\widehat{F}_{\ell,\omega}$ the main term of (\ref{feaf-cb-1}) can be made not greater than $\varepsilon[\widehat{F}_{\ell,\omega}(E)+o(\widehat{F}_{\ell,\omega}(E))]$ for large $E$ by appropriate choice of $\,t$. So, the  tightness of continuity bound (\ref{feaf-cb-1}) follows from (\ref{C-dif}),
since $\|\Phi-\Psi\|^{E}_{\diamond}\leq\|\Phi-\Psi\|_{\diamond}=2\,$ and $\,\lim_{E\rightarrow+\infty}(\widehat{F}_{\ell,\omega}(E)-F_{H_A}(E))=0$.

The above example also shows that the main term in  continuity bound (\ref{feaf-cb-2}) is close to the optimal one up to the factor $2$, since
$C(\Phi,H_A,E)=C_{\chi}(\Phi,H_A,E)$ and $C(\Psi,H_A,E)=C_{\chi}(\Psi,H_A,E)$ for the channels $\Phi$ and $\Psi$. $\square$ \medskip

The operational
definition of the entanglement-assisted classical capacity of an
infinite-dimensional energy-constrained quantum channel  is
given in~\cite{H-c-w-c,EAC}. By the  Bennett-Shor-Smolin-Thaplyal theorem adapted to constrained
channels (\cite[Theorem 1]{EAC}) the entanglement-assisted classical capacity
of an infinite-dimensional channel
$\,\Phi $ with the energy constraint (\ref{constraint}) is given by the
expression
\begin{equation*}
C_{\mathrm{ea}}(\Phi ,H_A,E)=\sup_{\Tr H_A\rho \leq E}I(\Phi,\rho),
\end{equation*}
where $I(\Phi,\rho)=I(B\!:\!R)_{\Phi\otimes\mathrm{Id}_{R}(\hat{\rho})}$,  $\hat{\rho}_{A}=\rho, \rank \hat{\rho}=1$,
is the quantum mutual information of the channel $\Phi$ at a state $\rho$.
Proposition \ref{omi} with $n=1$ implies the following \smallskip

\begin{property}\label{feaf-cb+} \emph{Let $\,\Phi$ and $\,\Psi$ be  quantum channels from $A$ to $B$ and  $\,\frac{1}{2}\shs \|\Phi-\Psi\|^{E}_{\diamond}\leq\varepsilon$, where $\|.\|^{E}_{\diamond}$ is the ECD-norm defined in (\ref{E-sn}).}

\smallskip

A) \emph{If the Hamiltonian $H_A$  satisfies condition (\ref{H-cond}) then
\begin{equation}\label{feaf-cb-3}
\begin{array}{rl}
|C_{\mathrm{ea}}(\Phi,H_A,E)-C_{\mathrm{ea}}(\Psi,H_A,E)|\, \leq & \!\! 2\varepsilon(2t+r_{\!\varepsilon}(t))\widehat{F}_{H_A}\!\!\left(\frac{E}{\varepsilon t}\right)\\\\
\,+ & \!\!2g(\varepsilon r_{\!\varepsilon}(t))+4h_2(\varepsilon t)
\end{array}
\end{equation}
for any $\,t\in(0,\frac{1}{2\varepsilon}]$, where $\,r_{\!\varepsilon}(t)=(1+t/2)/(1-\varepsilon t)$ and
$\widehat{F}_{H_A}(E)$ is any upper bound for the function $F_{H_A}(E)$ (defined in (\ref{F-def}))  with properties  (\ref{F-prop-1}) and (\ref{F-prop-2}), in particular, $\,\widehat{F}_{H_A}(E)=F_{H_A}(E+E^{A}_0)$.} \smallskip

\emph{If $\,A$ is the $\ell$-mode quantum oscillator then the right hand side of (\ref{feaf-cb-3})  can be rewritten as follows
\begin{equation*}
2\varepsilon(2t+r_{\!\varepsilon}(t))\!\left[\displaystyle
\widehat{F}_{\ell,\omega}(E)-\ell\log(\varepsilon t)\right]+2g(\varepsilon r_{\!\varepsilon}(t))+4h_2(\varepsilon t),
\end{equation*}
where $\widehat{F}_{\ell,\omega}(E)$ is defined in (\ref{bF-ub}). In this case continuity bound (\ref{feaf-cb-3}) with optimal $\,t$ is  tight for large $E$.}\medskip

B) \emph{If the channels $\Phi$ and $\Psi$ satisfy  condition (\ref{feaf}) and the Hamiltonian $H_B$  satisfies condition (\ref{H-cond}) then
\begin{equation}\label{feaf-cb-4}
\begin{array}{rl}
|C_{\mathrm{ea}}(\Phi,H_A,E)-C_{\mathrm{ea}}(\Psi,H_A,E)|\, \leq & \!\! 2\varepsilon(2t+r_{\!\varepsilon}(t))\widehat{F}_{H_B}\!\!\left(\frac{kE}{\varepsilon t}\right)\\\\
\,+ & \!\!2g(\varepsilon r_{\!\varepsilon}(t))+4h_2(\varepsilon t)
\end{array}
\end{equation}
for any $\,t\in(0,\frac{1}{2\varepsilon}]$, where $\widehat{F}_{H_B}(E)$ is any upper bound for the function $F_{H_B}(E)$ with properties  (\ref{F-prop-1}) and (\ref{F-prop-2}), in particular, $\,\widehat{F}_{H_B}(E)=F_{H_B}(E+E^{\!B}_0)$.}\smallskip

\emph{If $\,B$ is the $\ell$-mode quantum oscillator then the right hand side of (\ref{feaf-cb-4}) can be  rewritten as follows
\begin{equation*}
2\varepsilon(2t+r_{\!\varepsilon}(t))\!\left[\displaystyle
\widehat{F}_{\ell,\omega}(kE)-\ell\log(\varepsilon t)\right]+2g(\varepsilon r_{\!\varepsilon}(t))+4h_2(\varepsilon t),
\end{equation*}
where $\widehat{F}_{\ell,\omega}(E)$ is defined in (\ref{bF-ub}). In this case continuity bound (\ref{feaf-cb-4}) with optimal $\,t$ is  tight for large $E$.}\smallskip

\end{property}\medskip

Note that continuity bound (\ref{feaf-cb-3}) holds for \emph{arbitrary} channels $\Phi$ and $\Psi$. \smallskip

\emph{Proof.} A) Let $\H_R\cong\H_A$ and $H_R$ be an operator in
$\H_R$ unitarily equivalent to $H_A$. For any state $\rho$ satisfying the condition
$\Tr H_A\rho \leq E$ there exists a purification $\hat{\rho}\in\S(\H_{AR})$ such that
$\Tr H_R\hat{\rho}_{R}\leq E$. Since
\begin{equation*}
 I(\Phi,\rho)=I(B\!:\!R)_{\sigma}\quad \textrm{and} \quad I(\Psi,\rho)=I(B\!:\!R)_{\varsigma},
\end{equation*}
where $\sigma=\Phi\otimes\mathrm{Id}_{R}(\hat{\rho})$ and $\varsigma=\Psi\otimes\mathrm{Id}_{R}(\hat{\rho})$
are states in $\S(\H_{BR})$ such that $\Tr H_R\sigma_{R},\Tr H_R\varsigma_{R}\leq E$ and $\,\|\sigma-\varsigma\|_1\leq \|\Phi-\Psi\|^{E}_{\diamond}$, Proposition 2 in \cite{CHI} with trivial $C$ shows that the value of
$|I(\Phi,\rho)-I(\Psi,\rho)|$ is upper bounded by the right hand side of (\ref{feaf-cb-3}).\smallskip

B) Continuity bound (\ref{feaf-cb-4}) is obtained similarly, since in this case we have 
$\Tr H_B\sigma_{B},\Tr H_B\varsigma_{B}\leq kE$.\smallskip

The specifications concerning the cases when either $A$ or $B$ is the $\ell$-mode quantum oscillator follow
from the inequality $\,\widehat{F}_{\ell,\omega}(E/x)\leq \widehat{F}_{\ell,\omega}(E)-\ell\log x\,$ valid for any positive $E$ and $x\leq1$.

The tightness of the continuity bounds (\ref{feaf-cb-3}) and (\ref{feaf-cb-4}) can be shown assuming that $\Phi$ is the identity channel from the $\ell$-mode  quantum oscillator $A$ to $B=A$ and $\Psi$ is the  completely depolarizing channel with the vacuum output state.  It suffices to note that
$\,C_{\mathrm{ea}}(\Phi,H_A,E)=2F_{H_A}(E)\,$ and $\,C_{\mathrm{ea}}(\Psi,H_A,E)=0\;$ and to repeat the arguments from the proof of Proposition \ref{feaf-cb}. $\square$
\smallskip

If a function $\widehat{F}_{H}$ satisfies condition (\ref{F-prop-2}) then  $\,\displaystyle \lim_{x\rightarrow+0}x \widehat{F}_{H}(E/x)=0$. So,
Propositions \ref{SCT}, \ref{feaf-cb} and \ref{feaf-cb+} imply the following observations.

\smallskip

\begin{corollary}\label{qc-cont} \emph{Let $\,\mathfrak{F}(A,B)$ be the set of all quantum channels from $A$ to $B$ equipped with the strong convergence topology (described in Section 3).}

\smallskip

A) \emph{If the Hamiltonians $H_A$ and $H_B$ satisfy, respectively, the condition of Proposition \ref{SCT}B and condition (\ref{H-cond}) then the functions
$$
\Phi\mapsto C_{\chi}(\Phi,H_A,E),\quad  \Phi\mapsto C(\Phi,H_A,E)\quad\textit{and} \quad\Phi\mapsto C_{\mathrm{ea}}(\Phi,H_A,E)
$$
are uniformly continuous\footnote{see Remark \ref{CHI-CB-r}.} on any subset of
$\,\mathfrak{F}(A,B)$ consisting of channels with bounded energy amplification factor (for the given input energy $E$).}

\smallskip

B) \emph{If the Hamiltonian $H_A$ satisfies conditions (\ref{H-cond}) then the function\break
$\,\Phi\mapsto C_{\mathrm{ea}}(\Phi,H_A,E)$  is uniformly continuous on
$\,\mathfrak{F}(A,B)$.}
\end{corollary}

\bigskip

\textbf{Note Added:} After posting the first version of this paper I was informed by A.Winter that he and his colleagues independently have come to the same "energy bounded" modification of the diamond norm. I am grateful to A.Winter for sending me a draft of their paper \cite{W-EBN}, which complements the present paper by detailed study of the ECD-norm from the physical point of view and by continuity bounds for the quantum and private classical capacities of energy-constrained channels. I hope it will appear soon.

\bigskip

I am grateful to A.Winter for valuable communication, in particular, for his talk about drawbacks of the diamond-norm topology in infinite dimensions motivating this work. I am also grateful to A.S.Holevo and G.G.Amosov for useful discussion. Special thanks to A.V.Bulinski for essential suggestions used in preparing this paper. \smallskip

The research is funded by the grant of Russian Science Foundation
(project No 14-21-00162).

\end{document}